\documentclass{article}
\usepackage{arxiv}
\usepackage[utf8]{inputenc} 
\usepackage[T1]{fontenc}    
\usepackage{hyperref}       
\usepackage{url}            
\usepackage{booktabs}       
\usepackage{amsfonts}       
\usepackage{nicefrac}       
\usepackage{microtype}      
\usepackage{lipsum}		
\usepackage{graphicx}
\usepackage{mathtools}
\usepackage{doi}

\title{Computation of Reliability Statistics for Success-Failure Experiments}

\date{February 2, 2023}	

\author{ \href{https://orcid.org/0000-0002-9700-0749}{\includegraphics[scale=0.06]{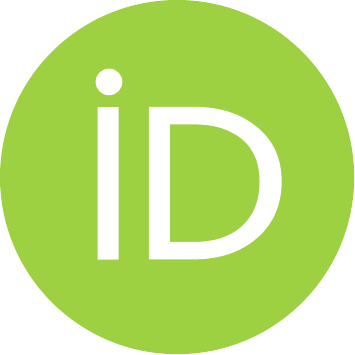}\hspace{1mm}Sanjay M.~Joshi}\thanks{https://www.linkedin.com/in/sanjaymjoshi/} \\
	Independent Researcher \\
	\texttt{sanjaymjoshi@iitbombay.org} \\
}

\renewcommand{\shorttitle}{Reliability Statistics for Bernoulli Trials}

\hypersetup{
pdftitle={Reliability Statistics for Bernoulli Trials},
pdfsubject={math.ST, stat.AP, stat.CO},
pdfauthor={Sanjay M.~Joshi},
pdfkeywords={Confidence, Reliability, Assurance, Sample Sizes, Bernoulli Trials},
}

\begin{document}
\maketitle

\begin{abstract}
Reliability is probability of success in a success-failure experiment. Confidence in reliability estimate improves with increasing number of samples. Assurance sets confidence level same as reliability to create one number for easier communication. Assuming binomial distribution for the samples, closed-form expression exists only for calculating confidence. The Wilson Score interval with continuity correction provides approximate closed-form expression for reliability. Brent's method was found to provide fast and accurate estimate for both reliability and assurance computations. Graphs and tables are provided for several number of samples. Two open-source \texttt{python} libraries are introduced for computing reliability, confidence, and assurance.
\end{abstract}


\section{Introduction}
Reliability of a part, functionality, or experiment is the probability that it is successful or does not fail. Such success or failure experiments are often called Bernoulli or binomial trials. The outcome or sample is a random variable with binomial distribution. We will assume that the samples are independent of each other and are identically distributed.

This article presents a review of related statistical parameters of confidence and assurance and presents computational results for several sample sizes.

\subsection{Reliability}
Reliability is the probability of success for a sample. If we encounter $f$ failures in $n$ samples, the reliability estimate can be calculated as $\hat{r} = (n-f)/n$.

This is only an estimate, since we have observed only $n$ samples out of the population, assumed to be infinite. The actual reliability for the population may be different than the estimate. Therefore, we need to qualify the reliability estimate with a level of confidence.

\subsection{Confidence}
For a Bernoulli trial with $n$ samples, the probability of observing exactly $k$ failures in $n$ samples with reliability $r$ is \cite{papoulis84}
\begin{equation}
   Pr(k; n,  r) = f(k) = \binom{n}{k} (1-r)^k r^{n-k},
\end{equation}
where $\binom{n}{k} = \frac{n!}{k!(n-k)!}$ is the binomial coefficient.


Confidence, $c$, is the probability that the actual reliability is at least $r$. If there are $f$ failures in $n$ samples, the computation is \cite{iso16269}
\begin{equation}
    c = 1 - \sum_{k=0}^f \binom{n}{k}  (1-r)^k r^{n-k},
    \label{C_f}
\end{equation}

With zero failures ($k=0$) in $n$ samples, the computation is straightforward and is often seen in the literature
\begin{equation}
    c = 1 - r^n
    \label{C_0}
\end{equation}

Thus, when we communicate a reliability level based on a finite number of samples, we should also quote the associated confidence level. Using these two numbers can get confusing for someone not familiar with the concept of confidence and reliability. A simpler concept of assurance is described next.

\subsection{Assurance}

The concept of assurance was introduced by Fulton \cite{luko97} by simply setting reliability equal to confidence. For example, assurance of 90\% means 90\% reliability with 90\% confidence. Intuitively, a high value of reliability with low level confidence does not make much sense. Likewise, having high confidence in a low reliability level does not add much value.
Thus, the concept of assurance is sufficient for easier communication using only one number.

Set assurance, $a = r = c$, in Eq. \ref{C_0} for zero failures in $n$ samples to get
\begin{equation}
    a = 1 - a^n
    \label{A_0}
\end{equation}
and for $f$ failures in $n$ samples, use Eq. \ref{C_f} to get
\begin{equation}
    a = 1 - \sum_{k=0}^f \binom{n}{k}  (1-a)^k a^{n-k}
    \label{A_f}
\end{equation}

These equations are not easy to solve for $n>$, since computing $a$ given $n$ and $f$ is not a closed-form expression. Let's review computation of all the above parameters in the next section.

\section{Computations}

\subsection{Reliability}

In the above section, we saw that the computation of $c$ given $r$ and $n$ involves only closed-form expression. Instead, suppose we are given $c$ and $n$ and need to find $r$. An example is when we want to know the 95\% confidence interval of reliability given zero failures in 10 samples.

The zero failure case is straightforward indeed. Rewriting Eq. \ref{C_0} gives us a closed-form expression
\begin{equation}
    r = (1 - c )^{1/n}
\end{equation}

With non-zero failures, though, there is no closed-loop formula corresponding to Eq. \ref{C_f}. Since the samples have binomial distribution, the problem of finding $r$ is the same as finding the lower bound of the confidence interval at target error rate $\alpha = 1-c$.

A good approximation is "Wilson Score Interval", especially at smaller values of $n$ \cite{wallis13}. The lower bound of the interval is computed as

\begin{equation}
  r = \frac{1}{~1+\frac{\,z^2\,}{n}~}\left( \hat r+\frac{\,z^2\,}{2n} \right)
~ - ~
\frac{z}{~1+\frac{z^2}{n}~}\sqrt{\frac{\,\hat r(1-\hat r)\,}{n}+\frac{\,z^2\,}{4n^2}~},
\end{equation}
where $\hat r = (n - f)/n$, the fraction of good samples in the total number of samples. The parameter $z$ is the $(1 - \alpha/2)$ quantile of a standard normal distribution corresponding to the target error rate $\alpha$. 

The 'continuity correction' modification \cite{wallis13} replaces $\hat r$  with $\max(\hat r - \frac{1}{2n}, 0)$ in the above equation. Before we look at the accuracy of these, let's review another option.

It is possible to compute reliability using numerical methods. Equation \ref{C_f} can be rewritten as
\begin{equation}
    c + \sum_{k=0}^f \binom{n}{k}  (1-r)^k r^{n-k} - 1 = 0
    \label{c_expr}
\end{equation}

Given $n$, $f$, and $c$, the left-hand side can be solved numerically for $r$.

These calculations were implemented using \texttt{Python} and have been published as an open-source library \cite{joshi23}. The performance was tested using the default compute environment of Google Colaboratory, a 2.0 GHz Intel Xeon as a \texttt{Jupyter} notebook with public access \cite{joshi23a}. The computations took only a fraction of seconds in this compute configuration.

For the numerical approach, Brent's optimization \cite{brent73}  method (function \texttt{brentq} in \texttt{SciPy} library) with a tolerance of 0.1\% was used.

To evaluate the quality of Wilson Score Interval approximation, $n$ was set to 40. The number of failed samples, $f$ was increased from 1 to 39. For each resulting value of $r$, the confidence $c$ was computed using Eq. \ref{C_0}. This value of $c$ was used to compute $\hat{r}$ using the two Wilson score methods and the difference $\hat{r} - r$ was calculated to get the accuracy.

\begin{figure}[ht]
	\centering
	\includegraphics[scale=0.75]{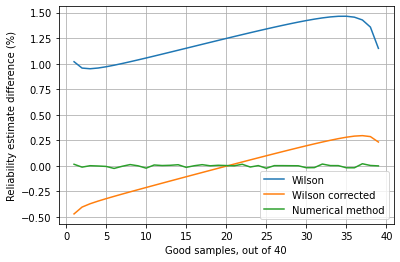}
	\caption{Accuracy of different methods of reliability computations}
	\label{fig:reli_closed}
\end{figure}

The accuracy plots for these methods are shown in Fig. \ref{fig:reli_closed}. As seen from the figure, the numerical method performs much better. 

\begin{figure}[ht]
	\centering
	\includegraphics[scale=0.75]{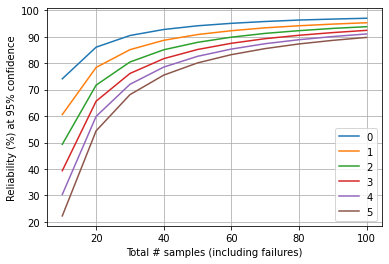}
	\caption{Reliability at 95\% confidence}
	\label{fig:reli_calc}
\end{figure}

To see how reliability changes at fixed confidence, the number of samples were changed from 10 to 100. The confidence was set to 95\% and the number of failures were increased from 0 to 5. The resulting plots using the numerical method are shown in Fig. \ref{fig:reli_calc} and the values are shown in Table \ref{tab:reli_95}

\begin{table}[htp]
    \centering
\begin{tabular}{r|rrrrrrrrrr}
\hline
    & \multicolumn{9}{c}{Total number of samples} \\
    &   10 &   20 &   30 &   40 &   50 &   60 &   70 &   80 &   90 &   100 \\
\hline
  0 & 74.1 & 86.1 & 90.5 & 92.8 & 94.2 & 95.1 & 95.8 & 96.3 & 96.7 &  97.0 \\
  1 & 60.6 & 78.4 & 85.1 & 88.7 & 90.9 & 92.3 & 93.4 & 94.2 & 94.8 &  95.3 \\
  2 & 49.3 & 71.7 & 80.5 & 85.1 & 87.9 & 89.9 & 91.3 & 92.3 & 93.2 &  93.8 \\
  3 & 39.3 & 65.6 & 76.1 & 81.7 & 85.2 & 87.6 & 89.3 & 90.6 & 91.6 &  92.5 \\
  4 & 30.3 & 59.9 & 72.0 & 78.6 & 82.6 & 85.4 & 87.4 & 88.9 & 90.1 &  91.1 \\
  5 & 22.2 & 54.4 & 68.1 & 75.5 & 80.1 & 83.3 & 85.5 & 87.3 & 88.7 &  89.8 \\
\hline
\end{tabular}
\caption{Reliability percentages for different number of failures at 95\% confidence}
\label{tab:reli_95}
\end{table}

\subsection{Confidence}

Equation \ref{C_f} is closed-form, and works well for small values of $n$ and $f$. As $n$ increases, especially above 1000, the calculation of $\binom{n}{k}$ becomes numerical unstable. To overcome this issue, let's rewrite Eq. \ref{C_f} as $c = 1- F(k)$, where $F(k)$ is the cumulative density function (CDF) of $f(k)$. 

The library in \cite{joshi23} uses the 'survival function' method of \texttt{binom} object from \texttt{SciPy} module for these computations.

Figure \ref{fig:reli_closed} shows how confidence levels change with number of failures $f$ and number of samples $n$ for a fixed level of reliability. The values are shown in Table \ref{tab:conf_90}.
\begin{figure}[ht]
	\centering
	\includegraphics[scale=0.75]{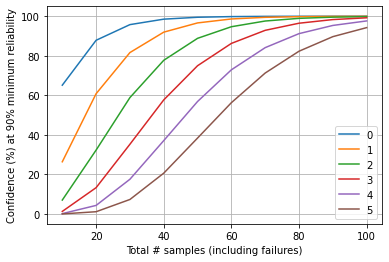}
	\caption{Confidence level with different number of failures}
	\label{fig:confidence_calc}
\end{figure}

\begin{table}[htp]
    \centering
\begin{tabular}{r|rrrrrrrrrr}
\hline
    & \multicolumn{9}{c}{Total number of samples} \\
    &   10 &   20 &   30 &   40 &   50 &   60 &   70 &    80 &    90 &   100 \\
\hline
  0 & 65.1 & 87.8 & 95.8 & 98.5 & 99.5 & 99.8 & 99.9 & 100.0 & 100.0 & 100.0 \\
  1 & 26.4 & 60.8 & 81.6 & 92.0 & 96.6 & 98.6 & 99.5 &  99.8 &  99.9 & 100.0 \\
  2 &  7.0 & 32.3 & 58.9 & 77.7 & 88.8 & 94.7 & 97.6 &  98.9 &  99.5 &  99.8 \\
  3 &  1.3 & 13.3 & 35.3 & 57.7 & 75.0 & 86.3 & 92.9 &  96.5 &  98.3 &  99.2 \\
  4 &  0.2 &  4.3 & 17.5 & 37.1 & 56.9 & 72.9 & 84.1 &  91.2 &  95.3 &  97.6 \\
  5 &  0.0 &  1.1 &  7.3 & 20.6 & 38.4 & 56.3 & 71.3 &  82.3 &  89.7 &  94.2 \\
\hline
\end{tabular}
\caption{Confidence percentages for different number of failures at 90\% reliability}
\label{tab:conf_90}
\end{table}

\subsection{Assurance}

To compute $a$ given $n$, we rewrite Eq. \ref{A_0} as
\begin{equation}
    a^n + a - 1 = 0
\end{equation}
Then the problem reduces to finding real root of the above equation. For a more generic version, rewrite Eq. \ref{A_f} as
\begin{equation}
    a + \sum_{k=0}^f \binom{n}{k}  (1-a)^k a^{n-k} - 1 = 0
\end{equation}
The library in \cite{joshi23} uses Brent's method for these computations.

\begin{figure}[ht]
	\centering
	\includegraphics[scale=0.75]{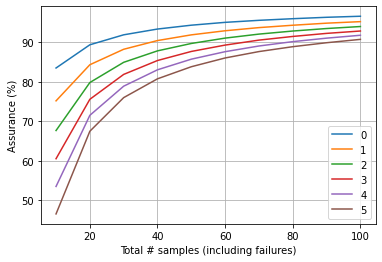}
	\caption{Assurance percentages with different number of failures}
	\label{fig:assurance_calc}
\end{figure}

Figure \ref{fig:assurance_calc} shows how assurance levels change with number of failures $f$ and number of samples $n$. The corresponding values are shown in Table \ref{tab:assurance}.

\begin{table}[tbp]
    \centering
    \begin{tabular}{r|rrrrrrrrrrrrrrr}
    \hline
    & \multicolumn{15}{c}{Total number of samples} \\
    &    1 &    2 &    3 &    4 &    5 &    6 &    7 &    8 &    9 &   10 &   11 &   12 &   13 &   14 &   15 \\
\hline
  0 & 50.0 & 61.8 & 68.2 & 72.4 & 75.5 & 77.8 & 79.7 & 81.2 & 82.4 & 83.5 & 84.5 & 85.2 & 86.0 & 86.6 & 87.2 \\
  1 &  0.0 & 38.2 & 50.0 & 57.2 & 62.3 & 66.1 & 69.1 & 71.5 & 73.5 & 75.2 & 76.7 & 77.9 & 79.1 & 80.0 & 81.0 \\
  2 &  0.0 &  0.0 & 31.8 & 42.8 & 50.0 & 55.3 & 59.4 & 62.7 & 65.4 & 67.7 & 69.6 & 71.3 & 72.9 & 74.2 & 75.4 \\
  3 &  0.0 &  0.0 &  0.0 & 27.6 & 37.7 & 44.7 & 50.0 & 54.2 & 57.6 & 60.5 & 63.0 & 65.1 & 66.9 & 68.6 & 70.1 \\
  4 &  0.0 &  0.0 &  0.0 &  0.0 & 24.5 & 33.9 & 40.6 & 45.8 & 50.0 & 53.5 & 56.4 & 59.0 & 61.2 & 63.2 & 65.0 \\
  5 &  0.0 &  0.0 &  0.0 &  0.0 &  0.0 & 22.2 & 31.0 & 37.3 & 42.4 & 46.5 & 50.0 & 53.0 & 55.6 & 57.9 & 59.9 \\
\hline
\end{tabular}
\begin{tabular}{r|rrrrrrrrrrrrrrr}
\hline
    & \multicolumn{15}{c}{Total number of samples} \\
    &   16 &   17 &   18 &   19 &   20 &   21 &   22 &   23 &   24 &   25 &   26 &   27 &   28 &   29 &   30 \\
\hline
  0 & 87.7 & 88.2 & 88.6 & 89.0 & 89.4 & 89.7 & 90.1 & 90.3 & 90.6 & 90.9 & 91.1 & 91.3 & 91.6 & 91.8 & 91.9 \\
  1 & 81.8 & 82.5 & 83.2 & 83.8 & 84.3 & 84.9 & 85.4 & 85.8 & 86.2 & 86.6 & 87.0 & 87.3 & 87.7 & 88.0 & 88.3 \\
  2 & 76.4 & 77.4 & 78.3 & 79.1 & 79.8 & 80.6 & 81.2 & 81.8 & 82.3 & 82.8 & 83.3 & 83.8 & 84.2 & 84.6 & 85.0 \\
  3 & 71.4 & 72.6 & 73.7 & 74.7 & 75.6 & 76.4 & 77.2 & 77.9 & 78.6 & 79.3 & 79.8 & 80.4 & 80.9 & 81.4 & 81.9 \\
  4 & 66.5 & 68.0 & 69.2 & 70.4 & 71.5 & 72.5 & 73.4 & 74.3 & 75.1 & 75.8 & 76.5 & 77.2 & 77.8 & 78.4 & 78.9 \\
  5 & 61.7 & 63.4 & 64.9 & 66.2 & 67.5 & 68.6 & 69.7 & 70.7 & 71.6 & 72.5 & 73.3 & 74.0 & 74.7 & 75.4 & 76.0 \\
\hline
\end{tabular}
\begin{tabular}{r|rrrrrrrrrrrrrrrr}
\hline
    & \multicolumn{15}{c}{Total number of samples} \\
    &   31 &   33 &   35 &   37 &   39 &   41 &   43 &   45 &   47 &   49 &   51 &   53 &   55 &   57 &   59 \\
\hline
  0 & 92.1 & 92.5 & 92.8 & 93.1 & 93.3 & 93.5 & 93.7 & 93.9 & 94.1 & 94.3 & 94.5 & 94.6 & 94.8 & 94.9 & 95.0 \\
  1 & 88.5 & 89.1 & 89.5 & 89.9 & 90.3 & 90.7 & 91.0 & 91.3 & 91.6 & 91.8 & 92.1 & 92.3 & 92.5 & 92.7 & 92.9 \\
  2 & 85.3 & 86.0 & 86.6 & 87.2 & 87.6 & 88.1 & 88.5 & 88.9 & 89.3 & 89.6 & 89.9 & 90.2 & 90.5 & 90.8 & 91.0 \\
  3 & 82.3 & 83.1 & 83.9 & 84.5 & 85.1 & 85.7 & 86.2 & 86.7 & 87.1 & 87.5 & 87.9 & 88.3 & 88.6 & 88.9 & 89.2 \\
  4 & 79.4 & 80.4 & 81.2 & 82.0 & 82.7 & 83.4 & 84.0 & 84.5 & 85.1 & 85.5 & 86.0 & 86.4 & 86.8 & 87.2 & 87.5 \\
  5 & 76.6 & 77.7 & 78.7 & 79.6 & 80.4 & 81.1 & 81.8 & 82.4 & 83.0 & 83.6 & 84.1 & 84.6 & 85.0 & 85.5 & 85.8 \\
\hline
\end{tabular}
\begin{tabular}{r|rrrrrrrrrrrrrrrr}
\hline
    & \multicolumn{15}{c}{Total number of samples} \\
    &   60 &   90 &   120 &   150 &   180 &   210 &   240 &   270 &   300 &   330 &   360 &   390 &   420 &   450 &   480 \\
\hline
  0 & 95.1 & 96.4 &  97.1 &  97.6 &  97.9 &  98.1 &  98.3 &  98.5 &  98.6 &  98.7 &  98.8 &  98.8 &  99.0 &  99.0 &  99.0 \\
  1 & 93.0 & 94.9 &  95.9 &  96.6 &  97.0 &  97.4 &  97.7 &  97.9 &  98.1 &  98.2 &  98.3 &  98.5 &  98.5 &  98.6 &  98.7 \\
  2 & 91.1 & 93.5 &  94.9 &  95.7 &  96.3 &  96.7 &  97.1 &  97.4 &  97.6 &  97.8 &  97.9 &  98.1 &  98.2 &  98.3 &  98.4 \\
  3 & 89.3 & 92.3 &  93.9 &  94.9 &  95.7 &  96.2 &  96.6 &  96.9 &  97.2 &  97.4 &  97.6 &  97.7 &  97.9 &  98.0 &  98.1 \\
  4 & 87.7 & 91.1 &  93.0 &  94.2 &  95.0 &  95.6 &  96.1 &  96.4 &  96.8 &  97.0 &  97.2 &  97.4 &  97.6 &  97.7 &  97.8 \\
  5 & 86.1 & 89.9 &  92.1 &  93.4 &  94.4 &  95.0 &  95.6 &  96.0 &  96.4 &  96.6 &  96.9 &  97.1 &  97.3 &  97.5 &  97.6 \\
\hline
\end{tabular}
\caption{Assurance percentages for different number of failures. The numbers are rounded. Assurance of 99\% is exceeded at 459 samples with zero failures.}
\label{tab:assurance}
\end{table}

\section{Conclusions}
We reviewed reliability as the probability of success of an experiment in success-failure experiments, where experiment outcomes have binomial distribution. When we calculate estimate of reliability based on a finite number of samples, we should qualify that estimate with level of confidence. We reviewed the concept of assurance that combines reliability and assurance into a single number for easier communication.

We looked at methods to compute reliability, confidence, and assurance.  Confidence can be calculated using a closed-form expression, but using survival function parameter of binomial distribution offers numerically stable values even at high number of samples. For computing reliability, the Wilson Score Interval with continuity correction may provide sufficient accuracy, typically less than 5\%, with a closed-form expression. Brent's method of numerically solving the equation yields more accurate estimate and is computationally fast. Computing assurance with high accuracy is feasible using Brent's method.

In this paper, we saw graphs and tables created from a \texttt{Jupyter} notebook using new open-source \texttt{python} library. Both of these can be easily modified to obtain statistics for reliability, confidence, and assurance.

\bibliographystyle{unsrtnat}


\end{document}